\documentclass[10pt,twocolumn,letterpaper]{revtex4}
\usepackage{graphicx}

\usepackage{amsmath}
\usepackage{amssymb}

\begin{document}


\title{Ultrafast Deflection of Spatial Solitons in Al$_{x}$Ga$_{1-x}$As
Slab Waveguides}
\author{J. H{\"u}bner and H. M. van Driel}
\affiliation{Department of Physics and Institute for Optical Sciences,
University of Toronto, 60 St. George Street, Toronto M5S 1A7, Canada}

\author{J. S. Aitchison}
\affiliation{Department of Electrical and Computer Engineering  and
Institute for Optical Sciences, University of Toronto, 10 King's College
Road, Toronto M5S 3G4, Canada}

\begin{abstract}
We demonstrate ultrafast all-optical deflection of spatial solitons in an
Al$_{x}$Ga$_{1-x}$As slab waveguide using 190\,fs, 1550\,nm pulses which
are used to generate and deflect the spatial soliton. The steering beam is
focused onto the top of the waveguide near the soliton pathway and the
soliton is steered due to refractive index changes induced by optical
Kerr, or free carrier (Drude) effects. Angular deflections up to 8\,mR are
observed.
\end{abstract}

\maketitle


\noindent Optical spatial solitons are shape-invariant wavepackets
maintained by the balancing of linear and nonlinear optical effects. The
nonlinear processes which compensate for diffraction and produce guiding
can be induced via, {\it e.g.}, an intensity dependent refractive index
(Kerr effect), photorefractive effects or cascaded second order optical
nonlinearities.\cite{TrilloSPRINGER} While interesting objects of study in
themselves, solitons are also being considered for information processing
applications such as optically reconfigurable logic devices
\cite{BlairAO1999}. The switching of spatial solitons is an essential
process for many applications, and popular embodiments include nonlinear
interactions of co-propagating spatial solitons \cite{ShiOL1990,kangOL96}
or electro-optically induced pathway distortions. All-optical ultrafast
reconfiguration of soliton pathways, e.g., due to the Kerr effect, in the
telecommunication wavelength regime (1.3-1.6\,$\mu$m) is an attractive
goal and offers higher switching speeds compared to electro-optic methods.
However, besides the co-propagating schemes one should consider other
techniques that have been discussed in the more general realm of ``light
by light{'}' switching such as optically or electro-optically induced
birefringence \cite{HaasAPLO64,ShwartzOL04}, optically induced prisms
\cite{LiOL1991} or gratings\cite{RoosenJAP54}.

\begin{figure}[tbp]
  \includegraphics[width=\columnwidth]{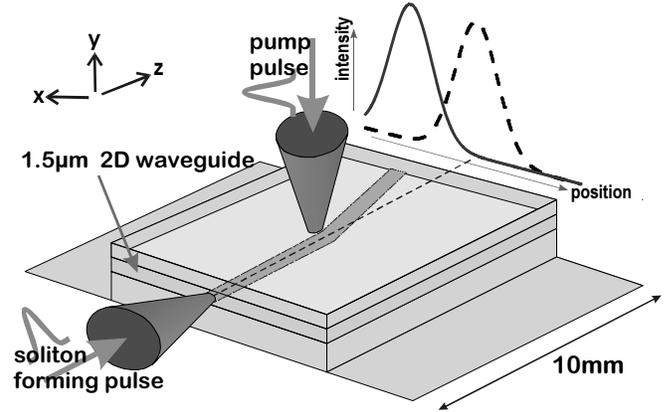}
  \caption{Principle of the soliton deflection technique: The spatial soliton
formed within the waveguide layer is deflected by a localized refractive
index change induced by a pump beam normally incident on the waveguide.}
\label{fig:principle}\vspace{-2mm}
\end{figure}

Here we demonstrate an ultrafast, non-planar switching scheme for spatial
solitons formed in a 2D Al$_{x}$Ga$_{1-x}$As waveguiding layer by inducing
a localized refractive index perturbation in the spatial soliton pathway
using femtosecond light pulses normally incident on the waveguide. The
general principle of the technique is depicted in
Fig.\,\ref{fig:principle} which shows an optical soliton formed at the
entrance facet of a 2D waveguide by an ultrashort laser pulse. A separate
ultrashort pump pulse is focussed onto the top of the waveguide,
introducing an index change $\Delta n$ in the waveguiding layer with the
spatial profile of the pump pulse. The $\Delta n$, which can be generated
on either side of the spatial soliton pathway, causes the soliton to
deflect while remaining intact: The robust nature of solitons propagation
means that it is not not necessary to form an optically induced prism of a
particular shape\cite{MamyshevEL1994}. The deflection direction also
depends on the sign of $\Delta n$ which is $>0$ for the optical Kerr
effect and $<0$ for any free carrier (Drude) induced index change in
Al$_{x}$Ga$_{1-x}$As. For $\Delta n>0$ the soliton path bends towards the
index gradient, whereas for $\Delta n<0$ deflection occurs in the opposite
direction. Temporal control of the deflection is achieved by delaying the
soliton forming pulse with respect to the pump pulse; temporal resolution
is related to the convolution of both pulses.

The waveguide structure we employ consists of a $1.5\,\mu$m thick, 10\,mm
long (in the $\langle 110\rangle$ direction) Al$_{x}$Ga$_{1-x}$As
waveguiding layer ($x=0.18$, band gap = 1.64\,eV $\equiv 756$ nm)
sandwiched in a $4\,\mu$m thick lower and a $1.5\,\mu$m upper cladding
both with $x=0.24$. The pulses used to generate and steer the solitons are
obtained from an ultrafast laser system: a 250\,kHz optical parametric
amplifier producing 190\,fs pulses, centered at $\lambda =1550$\,nm
(photon energy = 0.8\,eV). For soliton formation a 1.4\,kW peak power
pulse is elliptically shaped by a cylindrical telescope (to enable good
coupling into the waveguide) and subsequently passed through a 11\,mm
focal length lens to obtain an appropriate lateral width at the waveguide.
The in-plane intensity diameter is $\approx 28\,\mu$m (measured at
$e^{-1}$ points) at the entrance facet with the light polarized in-plane.
We observe that the soliton maintains its width for $>3\times$ the linear
diffraction length having a width of $\approx 80\,\mu$m at the exit facet.
The pump or steering pulse has a peak power of 220\,kW and is focused to a
diameter (at $e^{-1}$points) of $\approx 19\,\mu$m polarized parallel to
the soliton polarization. The peak intensity is a factor of five below the
damage threshold. For maximal deflection the center of the pump pulse
focus is located 0.5\,mm from the entrance facet and displaced laterally
from the soliton channel so that the soliton experiences the largest
lateral variation in the induced refractive index (see below). The time
delay, $\Delta t$, between the soliton and the pump pulse is controlled
with a delay stage.
\begin{figure}[h]
\center
  \includegraphics[width=0.95\columnwidth]{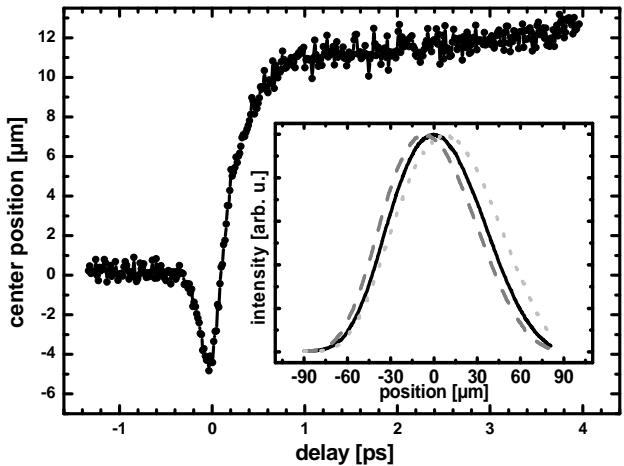}
  \caption{Center position of the soliton intensity distribution
  emerging from the exit facet versus the time delay, $\Delta t$, between the
soliton-forming and pump ultrashort laser pulses. The inset shows the
intensity profiles at the exit facet at $\Delta
t=-1$\,ps (solid curve), $\Delta t=0$ (dashed  curve) and $\Delta t=3$\,ps
(dotted curve). } \label{fig:ResultsOPAprofile} \vspace{-3mm}
\end{figure}

Fig.\,\ref{fig:ResultsOPAprofile} shows the change in the center of the
soliton lateral intensity distribution at the exit facet, as a function of
$\Delta t$. The fast, pulse width limited deflection towards the pump beam
position near $\Delta t=0$ is attributed to the third-order nonlinear
optical response, the optical Kerr effect, and peaks at $\approx 4.6\,\mu
$m. For an estimated peak focused intensity $I_{0}=75$\,GW\,cm$^{-2}$ and
a Kerr coefficient of\cite{AitchisonJQA1997} $n_{2}(\text{1550nm}) = 1.4
\times 10^{-4}\,$cm$^{2}$GW$^{-1}$ we estimate a peak refractive index
change of $n_{2}I_{0}=0.01$. However, for $\Delta t$ larger than the
temporal pulse width the deflection reverses, bends away from the pump
location and achieves a maximum value $>12\,\mu$m, due to the refractive
index change induced by free carriers. For even higher intensities,
obtained by tighter focusing of the pump beam, and yielding higher carrier
densities, deflections up to $80\,\mu$m have been achieved, corresponding
to a deflection angle of 8\,mR; this is larger than the soliton diameter
by more than a factor of two. The Drude-based change to the refractive
index via free carriers of density, $N$, is most likely due to
three-photon absorption; two-photon absorption is not energetically
allowed. Indeed, from Wherret's scaling
laws\cite{AitchisonJQA1997,WherrettJOSAB84} for multiphoton absorption we
estimate a 3-photon absorption coefficient of 0.05\,cm$^{3}$GW$^{-2}$, so
that the injected carrier density is estimated to be $N=3\times
10^{18}\text{cm}^{-3}$. From the refractive index change with carrier
density\cite{UlmerOL1999} $dn/dN\equiv\sigma_{n}=-7.4\times
10^{-21}\text{cm}^{3}$ for 1550\,nm, we obtain a peak refractive index
change of $\sigma_{n}N=0.02$.
\begin{figure}[h]
\center
  \includegraphics[width=0.95\columnwidth]{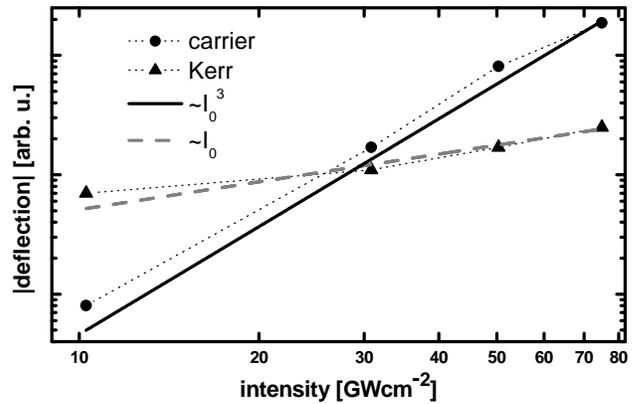} 
  \caption{Scaling of the Kerr- and carrier induced deflection magnitude
  measured with a position sensitive device.}
  \label{fig:PowerDep} \vspace{-3mm}
\end{figure}

Fig.\,\ref{fig:PowerDep} shows how the Kerr-induced and carrier-induced
deflection amplitudes scale with the focussed pump intensity $I_{0}$. As
expected the former scales linearly with $I_{0}$, consistent with a Kerr
effect, while the latter scales as $I_{0}^{3}$ in accord with three photon
absorption or a cascaded three-photon absorption processes. The
carrier-induced response rises on a time scale related to the pulse width,
although it might be noted that there is a ``slow{'}' (1 ps rise time)
component. This is attributed to cooling of the carriers, perhaps also
involving intervalley scattering, with the effective mass of conduction
band electrons decreasing and enhancing the Drude contribution to the
refractive index change. The carrier-induced response remains for the
carrier lifetime which is of the order of several hundreds picoseconds for
our sample. With appropriate engineering the recombination time could be
reduced or carrier induced changes could be enhanced. Alternatively
carrier responses can be avoided, by working with lower peak intensity by
e.g., using a larger focal area. The maximum deflection is then limited by
the pump pulse energy and the temporal pulse width. Note that with the
present experiments, the onset of the carrier-induced refractive index
change partly masks the Kerr effect and the actual Kerr induced deflection
in Fig.\,\ref{fig:ResultsOPAprofile} is likely to be much larger than the
$4.6\,\mu$m indicated.

A simple estimate of the observed deflection for refractive index changes
due to the optical Kerr effect $(\Delta n(I)=n_{2}I)$ can be given as
follows. Consider a planar slab waveguide contained along the y direction
with light propagating in the z direction as indicated in
Fig.\,\ref{fig:principle} The nonlinear wave equation is
\begin{equation}
\frac{\partial ^{2}A(x,z)}{\partial x^{2}}+\frac{n_{2}}{\eta _{0}}
k^{2}|A(x,z)|^{2}A(x,z)=2ik\frac{\partial A(x,z)}{\partial z}
\label{eq:NLSE}
\end{equation}
with $\eta_{0}=n_{0}c\epsilon_{0}/2\approx 4.3\,10^{-3}$ where $n_{0} =
3.24$ is the refractive index of the slab and $k=n_{0}2\pi /\lambda $. The
solution to (\ref{eq:NLSE}) is:
\begin{equation}
A(x,z)=A_{0}\,\text{sech}\left( \frac{x}{W_{0}}\right) \exp \left( -i\frac{z%
}{2kW_{0}^{2}}\right) .  \label{eq:solitonsolution}
\end{equation}
where $A_{0}=\sqrt{2\eta _{0}/n_{2}}/(kW_{0})$ is the amplitude of the
so-called bright soliton and $W_{0}$ is its width. A Kerr-induced change
in the refractive index along the x-direction, with the ultrafast pump and
soliton pulses overlapping in time, in addition to the one already induced
by the soliton itself, leads to a spatially dependent phase-shift for the
propagating soliton and therefore a refraction towards the direction of
the gradient of the phase-shift. The index variation induced by a pulse
focused onto the waveguide is taken to be of the form $\Delta
n(x,z)=n_{2}I(x,y)=n_{2}I_{0}\exp (-\left[ (x-x_{0})^{2}+z^{2}\right]
/w_{0}^{2})$ where $I_{0}$ is the peak intensity of the focused pump beam,
$w_{0}$ is its width and $(x_{0},\ z=0)$ is the center of the focal spot.
If the focused beam is located so that $\frac{\partial ^{2}I}{\partial
x^{2}}=0$ at the center of the soliton (i.e. for $x_{0}=w_{0}/\sqrt{2}$ )
and if we assume $w_{0}\gtrapprox 2W_{0}$, to a first approximation a
linear phase differential, $\phi (x,z),$ is induced in the soliton between
$x_{0}=\pm w_{0}/\sqrt{2}$ with:
\begin{equation}
\frac{\partial \phi (x,z)}{\partial x}\propto \frac{2\pi }{\lambda }\frac{%
\partial n}{\partial x}|_{x_{0}}=\frac{2\pi }{\lambda }\sqrt{\frac{2}{e}}%
\frac{n_{2}I_{0}}{w_{0}}e^{-\frac{(z-z_{0})^{2}}{w_{0}^{2}}}.
\label{eq:phaseshift}
\end{equation}
In general, a soliton travelling through the waveguide at a small angle
$\theta (\approx \sin \theta )$ with respect to the z-direction can be
described by\cite{HasegawaSPRINGER}: {\small
\begin{equation}
A(x,z)=A_{0}\,\text{sech}\left( \frac{x-\theta z}{W_{0}}\right) \exp
\left( -iz\frac{1-k^{2}kW_{0}^{2}\theta ^{2}}{2kW_{0}^{2}}-ixk\theta
\right) \label{eq:solitonsolutionangle}
\end{equation}
}which is equivalent to the solution in Eq.\,\ref{eq:solitonsolution} at
the place of deflection ($z=0$) multiplied by a phase factor $\exp
(-ixk\theta )$. The phase differential $\phi $ can be related to the
generic phase shift $xk\theta $ at $z=0$ through
\begin{eqnarray}
\int_{-W_{0}}^{W_{0}}k\theta \,dx &=&2W_{0}k\theta \equiv \\
\int_{-\infty }^{\infty }\frac{\partial \phi (x,z)}{\partial x}dz &=&\,\frac{%
\left( 2\pi \right) ^{3/2}}{\lambda e^{1/2}}\,n_{2}I_{0}w_{0}
\end{eqnarray}
This gives a deflection angle of
\begin{equation}
\theta \ \approx\ \frac{1}{2}\left( \frac{2\pi }{e}\right) ^{1/2}\frac{w_{0}%
}{W_{0}}\frac{n_{2}I_{0}}{n_{0}}  \label{eq:theta}
\end{equation}
With $w_{0}\approx 2\cdot W_{0}$ and $I_{0}\approx 75$\,GWcm$^{-2}$
Eq.\,\ref{eq:theta} gives $\theta \approx 2.5$\,mR. This is within a
factor of 3 the observed Kerr induced deflection angle of $\sim 1 $\,mR,
although the Drude effect partially obscures the Kerr effect. The
deflection angles reported here are comparable to those achieve with
switching techniques involving co-propagating solitons and electrically
induced prisms although the interaction lengths are much
longer\cite{kangOL96,friedrichOL98}.

In summary we have demonstrated the feasibility of all-optical alteration
of the pathways of spatial solitons in two dimensional semiconductor
waveguides on an ultrafast time scale, employing the optical Kerr
nonlinearity of the medium. The deflection scheme reported here does not
require a precise geometrical form for the refractive index change. It
also offers complementary degrees of freedom to the co-propagating soliton
switching scheme. In addition we have observed Drude-induced deflection of
spatial solitons. The ultrafast temporal nature and particular geometry of
the deflection technique enables one to write and perform information
processing applications using spatial solitons, with the maximum
deflection angle limited only by the maximum possible phase shift across
the soliton. By merging the capabilities of the approach described here
with the scheme of optical interconnects \cite{friedrichOL98} additional
applications are possible.

We acknowledge the support of the Sciences and Engineering Research
Council, Canada and Photonics Research Ontario.


\begin{thebibliography}{10}

\bibitem{TrilloSPRINGER}
S.~Trillo and W.~Toruellas, eds., \emph{Spatial Solitons}, vol.~82 of
  \emph{Springer Series in Optical Sciences} (Springer, 2001).

\bibitem{BlairAO1999}
S.~Blair and K.~Wagner, Appl. Opt. \textbf{38}(32), 6749 (1999).

\bibitem{ShiOL1990}
T.-T.~Shi and S.~Chi, Opt. Lett. \textbf{15}(20), 1123 (1990).

\bibitem{kangOL96}
J.~U. Kang, G.~I. Stegeman, and J.~S. Aitchison, Opt. Lett.
\textbf{21}(3), 189 (1996).

\bibitem{HaasAPLO64}
W.~Haas, R.~Johannes, and P.~Cholet, Appl. Opt. \textbf{3}(8), 988 (1964).

\bibitem{ShwartzOL04}
S.~Shwartz, M.~Segev and U.~El-Hannay, Opt. Lett. \textbf{29}(7), 760
(2004).

\bibitem{LiOL1991}
Y.~Li, D.~Y. Chen, L.~Yang, and R.~R. Alfano, Opt. Lett. \textbf{16}(6),
438 (1991).

\bibitem{RoosenJAP54}
G.~Roosen and G.~T. Sincerbox, J. Appl. Phys. \textbf{54}(3), 1628 (1983).

\bibitem{MamyshevEL1994}
P.~V. Mamyshev, A.~Villeneuve, G.~I. Stegeman, and J.~S. Aitchison,
Electron. Lett. \textbf{30}(9), 726 (1994).

\bibitem{AitchisonJQA1997}
J.~S. Aitchison, D.~C. Hutchings, J.~U. Kang, G.~I. Stegeman, and
  A.~Villeneuve, IEEE JOURNAL OF QUANTUM ELECTRONICS \textbf{33}(3), 341
  (1997).

\bibitem{WherrettJOSAB84}
B.~S. Wherrett, J. Opt. Soc. Amer.~B \textbf{1}, 67 (1984).

\bibitem{UlmerOL1999}
T.~G. Ulmer, R.~K. Tan, Z.~Zhou, S.~E. Ralph, R.~P. Kenan, C.~M. Verber,
and J.~SpringThorpe, Opt. Lett. \textbf{24}(11), 756 (1999).

\bibitem{HasegawaSPRINGER}
A.~Hasegawa and M.~Matsumoto, \emph{Optical Solitons in Fibers} (Springer,
  2003).

\bibitem{friedrichOL98}
L.~Friedrich, G.~I. Stegeman, P.~Millar, C.~J. Hamilton, and J.~S.
Aitchison, Opt. Lett. \textbf{23}(18), 1438 (1998).

\end{thebibliography}
\end{document}